\documentclass[a4paper,prx,twocolumn,superscriptaddress,amsmath,amssymb,mathrsfs]{revtex4}
\usepackage{tabularx}
\usepackage{graphicx}
\usepackage{braket}
\usepackage{dcolumn}
\usepackage{bm}
\usepackage{epsfig}
\usepackage{rotating}
\usepackage{color}
\usepackage{hyperref}
\hypersetup{colorlinks=true,linkcolor=blue,citecolor=blue,urlcolor=blue}

\begin{document}

\newcommand{\minusone}{{-\!1}}

\title{Lie algebra for rotational subsystems of a driven asymmetric top}

\author{E. Pozzoli}
\thanks{These authors have contributed equally.}
\affiliation{	Laboratoire Jacques-Louis Lions, Sorbonne Universit\'e, Universit\'e de Paris, CNRS, Inria, Paris, France}

\author{M. Leibscher}
\thanks{These authors have contributed equally.}
\affiliation{Theoretische Physik, Universit\"{a}t Kassel,
Heinrich-Plett-Stra{\ss}e 40, 34132 Kassel, Germany}

\author{M. Sigalotti} 
\affiliation{Laboratoire Jacques-Louis Lions, Sorbonne Universit\'e, Universit\'e de Paris, CNRS, Inria, Paris, France}

\author{U. Boscain} 
\affiliation{Laboratoire Jacques-Louis Lions, Sorbonne Universit\'e, Universit\'e de Paris, CNRS, Inria, Paris, France}

\author{C. P. Koch} 
\email{christiane.koch@fu-berlin.de}
\affiliation{Theoretische Physik, Universit\"{a}t Kassel,
Heinrich-Plett-Stra{\ss}e 40, 34132 Kassel, Germany}
\affiliation{Dahlem Center for Complex Quantum Systems and Fachbereich Physik, Freie Universit\"{a}t Berlin,
Arnimallee 14, 14195 Berlin, Germany}

\date{\today}
 
\begin{abstract}
  We present an analytical approach to construct the Lie algebra of finite-dimensional subsystems of the driven asymmetric top rotor. Each rotational level is degenerate due to the isotropy of space, and the degeneracy increases with rotational excitation. For a given rotational excitation, we determine the nested commutators between drift and drive Hamiltonians using a graph representation. We then generate the Lie algebra for subsystems with arbitrary rotational excitation using an inductive argument.
\end{abstract}

\maketitle

\section{Introduction}

Lie algebras, encoding the structure of Lie groups, are an essential tool to study symmetries in physics~\cite{GilmoreBook}.
\textit{Dynamical} Lie algebras characterize the coherent dynamics and symmetry behavior of a quantum system and thus play a central role in quantum control~\cite{Alessandro2008,GlaserEPJD15}. Given the Hamiltonian of a quantum system, its dynamical Lie algebra is constructed by taking the nested commutators of the drift, i.e., the field-free term, and all drives, i.e., all couplings to external fields. Since the Lie algebra elements are the generators of the dynamics, any time evolution can -- in principle (i.e., upon suitable choice of external fields) -- be generated if the dynamical Lie algebra is of full rank~\cite{Alessandro2008}.
For the simplest example of a two-level system, two non-commuting terms in the Hamiltonian, for example a $\sigma_z$-drift and $\sigma_x$-drive, are sufficient for the corresponding Lie algebra to be of full rank. In contrast, a $\sigma_z$-drive would not be enough to transfer the two-level system from any initial into any final state.

The construction of the dynamical Lie algebra quickly becomes challenging as the dimension of Hilbert space increases. 
For composite quantum systems such as $N$ two-level systems, one may start from the Lie algebras of the constituent systems but the presence or absence of interactions, i.e., entangling operations, renders the problem non-trivial~\cite{Alessandro2008}. For large Hilbert spaces that cannot be written as a tensor product, 
few methods exist to generate the elements of the Lie algebra and often one needs to resort to numerical approaches~\cite{Schirmer}. Such large Hilbert spaces may, however, display a tensor sum structure. This suggests to first construct the Lie algebra in a small subspace and then infer the elements in other subspaces.

Here, we show how to systemize this approach and construct the dynamical Lie algebra of a resonantly driven asymmetric top rotor in arbitrarily large rotational subspaces.
The driven quantum asymmetric top is an important paradigm of molecular physics with current applications ranging from quantum information~\cite{AlbertPRX20} to high-resolution spectroscopy~\cite{DomingosAnnuRevPhysChem18}.
Isotropy of space makes a quantum rotor an inherently degenerate system.
Orientational degeneracy is a challenge for quantum control since selecting a particular rotational state cannot be achieved by spectral selection alone~\cite{ShapiroBook}. However, the symmetry that is at the core of the degeneracy also provides the intuition for how to make a quantum rotor controllable --- by choosing drives, i.e., polarization directions, that break the symmetry. This was first realized for linear rotors~\cite{Judson1990}, where an inductive argument was used to prove approximate controllability and reachability of any state in a finite-dimensional subspace of the rotational spectrum at zero rotational temperature. A theory to decouple a finite-dimensional subspace from the rest of an infinite-dimensional spectrum~\cite{chambrion,CMSB,MR3101605,BCCS} allows to rigorously extend the 
proof of controllability of a linear rotor to unitary evolutions~\cite{BCS}. The controllability results of Ref.~\cite{BCS} have been recently generalized to symmetric top rotors in \cite{Boscain19}.
In comparison to linear and symmetric quantum rotors, asymmetric tops have a far more complex energy level structure. The conditions for unitary evolution controllability have only recently been identified for three-level subsystems $J/J+1/J+1$ with rotational quantum numbers $J=0$ and $J=1$~\cite{Leibscher20}. Here, we extend the proof to arbitrary rotational excitation. This is made possible by representing the Hamiltonian on a graph before making use of an inductive argument to determine the nested commutators generating the Lie algebra.

The paper is organized as follows. We briefly recall the model for an asymmetric quantum top in Sec.~\ref{sec:model} and introduce the graph representation in Sec.~\ref{sec:preliminaries}. The induction is carried out in Sec.~\ref{sec:induction}, and Sec.~\ref{sec:concl} concludes.

\section{Driven asymmetric top rotor}
\label{sec:model}

\begin{figure}[tb]
	\includegraphics[width=1.0\linewidth]{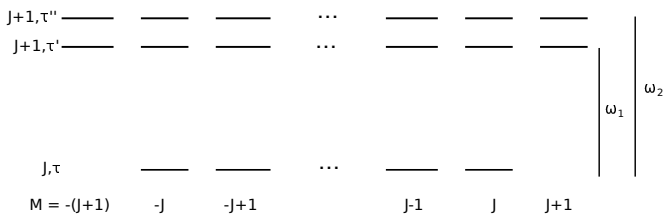} 
	\caption{Rotational subsystem consisting of the rotational states
		$|J,\tau,M \rangle$, $|J+1,\tau',M \rangle$ and $|J+1,\tau'',M \rangle$.}
	\label{model}
\end{figure}

We consider a subsystem of the spectrum of an asymmetric top as shown in Fig.\ref{model}, where the bars indicate the eigenstates $|J,\tau,M\rangle$ of the 
asymmetric top Hamiltonian
\begin{equation}
\hat H_0 =  A {\hat J_a}^2 + B {\hat J_b}^2 + C {\hat J_c}^2\,,
\label{eq:hrot}
\end{equation}
where $\hat J_a$, $\hat J_b$, and $\hat J_c$ are the angular momentum operators with respect to the principal molecular axes,
and $A > B > C$ are the rotational constants.

Here $J=0,1,2,...$ is the rotational quantum number. For each $J$ there exist $2J+1$ different rotational energy levels $E_{J,\tau}$, denoted by $\tau=-J,-J+1,...,J$. Each of those levels is $(2J+1)$-fold degenerate, with the degenerate stats denoted by the orientational quantum number $M=-J,-J+1,...,J$. 
Rotational subsystems of this kind are relevant for example for the enantiomer-selective excitation of rotational states of chiral molecules using microwave three-wave mixing~\cite{EibenbergerPRL17,PerezAngewandte17,PerezJPCL18}. The Hilbert space  $\mathcal{H}$ of the rotational subsystem is
\begin{widetext}
	\begin{eqnarray*}
		\mathcal{H}=\mathrm{span}\{|J,\tau,M\rangle\mid M=-J,\dots, J\} \oplus \mathrm{span}\{|J+1,\tau', M\rangle,|J+1,\tau'',M\rangle 
		\mid M=-(J+1),\dots,(J+1)\,\},  
	\end{eqnarray*}
\end{widetext}
with $\mathcal{H} \cong \mathbb{C}^{6J+7}$. Written in the basis of eigenstates $|J,\tau,M\rangle$, the rotational Hamiltonian $\hat {H}_{0}$  is given by a diagonal $(6J+7) \times (6J+7)$-matrix containing  $(2J+1)$-times the entry $E_{J,\tau}$, $(2J+3)$-times $E_{J+1,\tau'}$, and $(2J+3)$-times $E_{J+1,\tau''}$. The corresponding matrix is denoted by ${\bf H}_0$.
The asymmetric top interacts, via dipole interaction $ -{\hat {\vec \mu}} {\vec E}_i(t)$, with a set of $f$ electromagnetic fields such that
\begin{equation}
\hat H(t) = \hat H_0 - \sum_{i=1}^f \hat \mu_i E_i(t) = \hat H_0 + \sum_{i=1}^{f} \hat H_i E_i (t) \,,
\label{interaction}
\end{equation}
where where the electric fields is denoted by
${\vec E}_i(t) = {\vec e}_i E_i(t)$ with polarization vector ${\vec e}_i$ (equal to  either ${\vec e}_x$, ${\vec e}_y$, or ${\vec e}_z $) and amplitude
$E_i(t) = {\cal E}_i(t) \cos (\omega_i t + \varphi_i)$ 
where ${\cal E}_i(t)$ is the envelope and $\omega_i$ and $\varphi_i$ are  frequency and phase of the field. The dipole moment ${\hat {\vec \mu}}$ with the Cartesian components $\hat \mu_i$ equal to $\hat \mu_x$, $\hat \mu_y$ or $\hat \mu_z$
is given in the laboratory-fixed frame. Transformation to the to the dipole moments $\mu_\sigma =( \mu_a ,  \mu_b ,  \mu_c  )$ in the molecule-fixed frame by a rotation~\cite{Zare88} results in
\begin{widetext}
	\begin{eqnarray}
	\hat\mu_x   &=& \frac{\mu_a }{\sqrt{2}}  \left( D_{-10}^1  - D_{10}^1 \right)  + \frac{\mu_b }{ 2} \left ( D_{11}^1 - D_{1-1}^1  - D_{-11}^1 + D_{-1-1}^1\right) -  \mathrm{i} \frac{\mu_c }{2} \left ( D_{11}^1 + D_{1-1}^1  - D_{-11}^1 - D_{-1-1}^1\right), \nonumber \\
	\hat\mu_y   &=& -\mathrm{i} \frac{\mu_a }{\sqrt{2}}  \left( D_{-10}^1  + D_{10}^1 \right)  + \mathrm{i} \frac{\mu_b }{ 2} \left ( D_{11}^1 - D_{1-1}^1  + D_{-11}^1 - D_{-1-1}^1\right) + \frac{\mu_c }{2} \left ( D_{11}^1 + D_{1-1}^1  + D_{-11}^1 + D_{-1-1}^1\right), \nonumber \\
	\hat \mu_z   &=&  \mu_a  D_{00}^1 - \frac{\mu_b }{\sqrt{2} } \left ( D_{01}^1 - D_{0-1}^1 \right) +  \mathrm{i} \frac{\mu_c }{\sqrt{2}} \left ( D_{01}^1 + D_{0-1}^1 \right),
	\label{mu_projection}
	\end{eqnarray}  
\end{widetext}
where $D_{MK}^J$ denote the elements of the Wigner $D$-matrix. To evaluate the the matrix elements of the interaction Hamiltonian ${\hat H}_i$ in the asymmetric top basis, the asymmetric top eigenstates are written as a superposition of symmetric top eigenstates 
$|J,K,M\rangle$ \cite{KochRMP},
\begin{equation}
| J, \tau, M \rangle = \sum_{K} c_{K}^{J} (\tau) |J,K,M\rangle  \,,
\label{asym_top}
\end{equation}
with quantum number $K=-J,-J+1,...,J$, where states with different $K$ but the same $J$ and $M$ are mixed. The matrix elements $\langle J'', \tau'', M'' | {\hat H}_i | J', \tau', M' \rangle$ thus contain expressions of the form
 \begin{widetext}
	\begin{eqnarray}
	\langle J'', \tau'', M'' | D_{MK}^1 | J', \tau', M' \rangle = 
	\sum_{K',K''} 
	c_{K'}^{J'} \left ( c_{K''}^{J''} \right )^\ast \langle J'', K'', M'' |D_{MK}^1|J', K', M' \rangle
	\label{transition_asym}
	\end{eqnarray}
with \cite{Zare88}
	\begin{eqnarray}
	\langle J'', K'', M'' | D^1_{MK} | J', K', M' \rangle &=& \sqrt{2 J'' +1} \sqrt{2 J' +1} (-1)^{M''+K''}  
	 \left(    \begin{array}{ccc}  J' & 1& J'' \\ M' & M & -M'' \end{array} \right)
	\left( \begin{array}{ccc} J' & 1 & J'' \\ K' & K & -K'' \end{array} \right).
	\label{w3j}
	\end{eqnarray}
\end{widetext}
The Wigner 3j-symbols in Eq.~\eqref{w3j} determine the selection rules, namely $J''-J' = 0, \pm 1$ and $K'' = K' + K$ as well as $M'' = M' + M$ where the value of $M$ in Eq.~(\ref{transition_asym}) is determined by the electric field polarization,
with $M =0$ for $z$-polarized and  $M = \pm 1$ for $x$- or $y$-polarized fields.
 The frequencies of the electric fields in Eq.(\ref{interaction}) are chosen to be resonant to one of the rotational transitions, i.e. either $\omega_1=|E_{J+1,\tau'}-E_{J,\tau}|$, $\omega_2=|E_{J+1,\tau''}-E_{J,\tau}|$ or $\omega_3=|E_{J+1,\tau''}-E_{J+1,\tau'}|$. The field intensity can then be chosen such that only those transitions resonant to the frequency of the field are excited~\cite{Leibscher20}. The interaction Hamiltonian is thus determined by the polarization direction $p=x,y,z$ of the corresponding field and its frequency $\omega_i$. We thus denote the interaction Hamiltonian in the asymmetric top eigenbasis as ${\bf H}_i={\bf H}_{\omega_i,p}$.

\section{Graph representation}
\label{sec:preliminaries}

In the following, we consider a set of four interaction operators, namely
\begin{equation}
\label{eq:chi_chiral}
  {\bf H}_{\omega_1,p_1}, {\bf H}_{\omega_1,p_2}, {\bf H}_{\omega_2,p_3}, {\bf H}_{\omega_2,p_4}\,,
\end{equation}
where $p_i$, $i=1-4$ can be any polarization direction, $x$, $y$, or $z$ as long as  the pairs $p_1$, $p_2$ and $p_3$, $p_4$ are not
the same and all three polarization directions $x$, $y$, $z$ are
present.  The corresponding control fields have the frequencies $\omega_1=|E_{J+1,\tau'}-E_{J,\tau}|$, and $\omega_2=|E_{J+1,\tau''}-E_{J,\tau}|$, as indicated in Fig.~\ref{model}.
It has been demonstrated in \cite{Leibscher20} that the rotational dynamics of a rotational subsystem as in Fig.~\ref{model} is controllable with this set of interaction operators for the case $J=0$. In order to extend this proof to a subsystem with arbitrary $J$ with Hilbert space $\mathcal{H} \cong \mathbb{C}^{6J+7}$, it is necessary to show that the Lie algebra is 
\begin{eqnarray}\label{su(n)}
L &:=& \mathrm{Lie}\{ \mathrm{i}{\bf H}_{0}, \mathrm{i}{\bf H}_{\omega_1,p_1},\mathrm{i}{\bf H}_{\omega_1,p_2}, \mathrm{i}{\bf H}_{\omega_2,p_3}, \mathrm{i}{\bf H}_{\omega_2,p_4} \} \nonumber \\
&=&\mathfrak{su}(6J+7).
\end{eqnarray} 
 A basis of the Lie algebra $\mathfrak{su}(n)$ are the generalized Paul matrices \cite{Boscain19},
\begin{eqnarray}\label{basis}
{\bf G}_{j,k}&=&e_{j,k}-e_{k,j}\,, \nonumber \\
{\bf F}_{j,k}&=&\mathrm{i}e_{j,k}+\mathrm{i}e_{k,j}\,, \nonumber \\
{\bf D}_{j,k}&=&\mathrm{i}e_{j,j}-\mathrm{i}e_{k,k}\,,
\end{eqnarray}
for $j,k=1,\ldots,n$. Here $e_{j,k}$ is the matrix whose entries are all zero except for the entry in row $j$ and column $k$ which is equal to one. For the rotational subsystem in Fig.~\ref{model}, $i,j=1,...,6J+7$. In order to prove Eq.~\eqref{su(n)}, we thus need to show that repeatedly taking commutators between  $i {\bf H}_{\omega_i,p}$ and  $i {\bf H}_0$ yields elements of the Lie algebra which are proportional to each of the operators ${\bf G}_{j,k}$, ${\bf F}_{j,k}$, and  ${\bf D}_{j,k}$ alone.
For these computations, we will exploit the following properties of the generalized Paul matrices: Their commutator relations read  
\begin{subequations}\label{eq:commutators}
\begin{eqnarray}
\left[ {\bf G}_{j,k},{\bf G}_{k,n} \right]&=& {\bf G}_{j,n}\,, \nonumber \\
\left[ {\bf F}_{j,k},{\bf F}_{k,n} \right] &=& -{\bf G}_{j,n}\,, \nonumber \\
\left[ {\bf G}_{j,k},{\bf F}_{k,n} \right] &=& {\bf F}_{j,n} \,, 
\label{relation1} 
\end{eqnarray}
and 
\begin{eqnarray}
\label{relation2}
\left[{\bf G}_{j,k},{\bf F}_{j,k}\right] &=&2{\bf D}_{j,k} \,,\nonumber \\ 
\left[{\bf F}_{j,k},{\bf D}_{j,k}\right] &=& 2{\bf G}_{j,k}\,.
\end{eqnarray}
Operators which couple disjunct pairs of states commute,   
\begin{equation}\label{relation3}
[{\bf T}_{j,k},{\bf U}_{j',k'}]=0 \quad \text{ if } \{j,k\}\cap  \{j',k'\}=\emptyset ,
\end{equation}
with ${\bf T},{\bf U}\in \{{\bf G},{\bf F},{\bf D}\}$.
Finally, the commutators with the rotational Hamiltonian are given by
\begin{eqnarray}\label{relation4}
\left[\mathrm{i}{\bf H}_0,{\bf G}_{j,k} \right] &=&-\Delta E_{k,j} {\bf F}_{j,k}\,, \nonumber \\ 
\left[\mathrm{i} {\bf H}_0,{\bf F}_{j,k}\right] &=& \Delta E_{k,j} {\bf G}_{j,k}\,.
\end{eqnarray}
\end{subequations}
where $\Delta E_{k,j}$ is the energy level spacing between states $j$ and $k$.

\begin{figure}[tb]
	\includegraphics[width=1.0\linewidth]{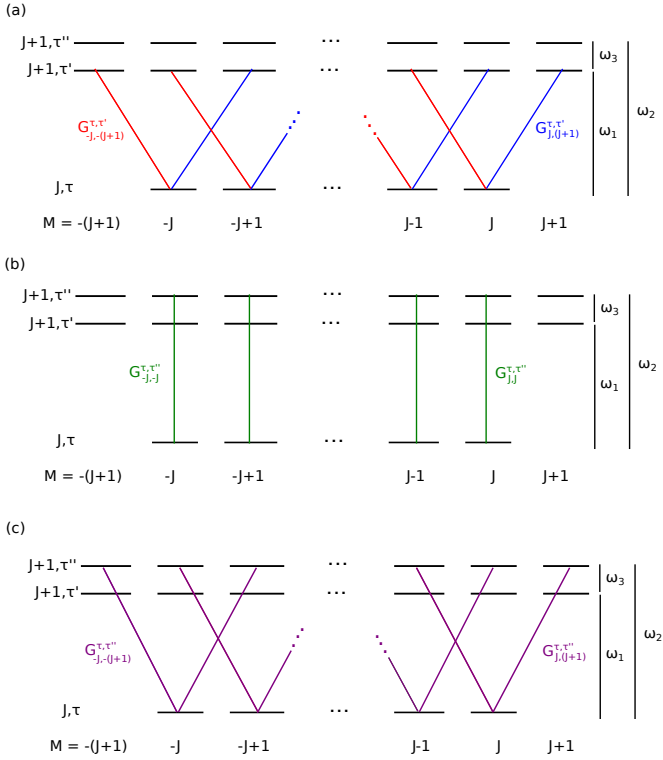} 
	\caption{
    (a) The red and blue lines indicate the transitions induced by the interaction Hamiltonians $\mathrm{i}{\bf H}_{\omega_1,x}$
		and $\mathrm{i}{\bf H}_{\omega_1,y}$. The blue (red) lines alone represent the interaction Hamiltonians $\mathrm{i}{\bf H}_{\omega_1,\sigma_+}$ ($\mathrm{i}{\bf H}_{\omega_1,\sigma_-}$). The green and purple lines present $\mathrm{i}{\bf H}_{\omega_2,z}$ (b) and
		$\mathrm{i}{\bf H}_{\omega_2,y}$ (c). Each of these lines represents one of the generalized Pauli matrices $G^{\tau,\tau'}_{M,M'}$.}
	\label{scheme_proof_general}
\end{figure}
To carry out the proof, we chose $p_1=x$, $p_2=y$, $p_3=y$ and $p_4=z$ and express the
anti-Hermitian operators $\mathrm{i}{\bf H}_{\omega_i,p}$ in terms of the generalized Pauli matrices.
Note that the coefficients $c_K^J$ in Eq.~\eqref{transition_asym} do not depend on $M$. The summation over these coefficients thus only results in a common prefactor, which is not relevant for the proof of controllability (i.e., for the generated Lie algebra) and can be factored out. 
For simplicity of notation, we denote the interaction Hamiltonians below 
without these $M$-independent prefactors. 
Note further that the $M$-dependence of the interaction Hamiltonians is solely determined by the $M$-dependent Wigner 3j-symbol in Eq.~\eqref{w3j}. For
$\mathrm{i}{\bf H}_{\omega_1,x}$, $\mathrm{i}{\bf H}_{\omega_1,y}$ and $\mathrm{i}{\bf H}_{\omega_2,y}$ in particular, it  is given by~\cite{Abramowitz64}
\begin{eqnarray*}
	\left(    \begin{array}{ccc}  J & 1& J+1 \\ M & \pm 1 & -(M \pm1) \end{array} \right) &=&	
	(-1)^{J+M}  \\ &\times& \frac{\sqrt{(J\pm M+2)(J \pm M+1)}}{\sqrt{(2J+3)(2J+2)(2J+1)}}\,,
\end{eqnarray*}
while for $\mathrm{i}{\bf H}_{\omega_2,z}$ it reads
\begin{eqnarray*}
	\left(    \begin{array}{ccc}  J & 1& J+1 \\ M & 0 & -M \end{array} \right) &=&	
	(-1)^{J+M}  \\ &\times& \frac{\sqrt{(J+M+1)(J-M+1)}}{\sqrt{(2J+3)(2J+1)(J+1)}}\,.
\end{eqnarray*}
We can thus write
\begin{widetext}
	\begin{eqnarray}
	\mathrm{i}{\bf H}_{\omega_1,x}&=&\mu_c\Big(\sqrt{(J+1)(2J+1)}({\bf G}^{\tau,\tau'}_{-J,-J-1}+{\bf G}^{\tau,\tau'}_{J,J+1})+\sqrt{J(2J+1)}({\bf G}^{\tau,\tau'}_{-J+1,-J}+{\bf G}^{\tau,\tau'}_{J-1,J}) \nonumber \\
	&+&...+ \sqrt{3} ({\bf G}^{\tau,\tau'}_{J-1,J-2}+{\bf G}^{\tau,\tau'}_{-J+1,-J+2})
	+({\bf G}^{\tau,\tau'}_{J,J-1}+{\bf G}^{\tau,\tau'}_{-J,-J+1})\Big),\label{ap_Hw1x}
	\\
	\mathrm{i}{\bf H}_{\omega_1,y}&=&\mu_c\Big(\sqrt{(J+1)(2J+1)}(-{\bf F}^{\tau,\tau'}_{-J,-(J+1)}+{\bf F}^{\tau,\tau'}_{J,J+1})+\sqrt{J(2J+1)}(-{\bf F}^{\tau,\tau'}_{-J+1,-J}+{\bf F}^{\tau,\tau'}_{J-1,J}) \nonumber \\
	&+&...+ \sqrt{3} (-{\bf F}^{\tau,\tau'}_{J-1,J-2}+{\bf F}^{\tau,\tau'}_{-J+1,-J+2})
	+(-{\bf F}^{\tau,\tau'}_{J,J-1}+{\bf F}^{\tau,\tau'}_{-J,-J+1})\Big),
	\label{ap_Hw1y} \\
	\mathrm{i}{\bf H}_{\omega_2,y}&=&\mu_a\Big(\sqrt{(J+1)(2J+1)}(-{\bf G}^{\tau,\tau''}_{-J,-(J+1)}+{\bf G}^{\tau,\tau''}_{J,J+1})+\sqrt{J(2J+1)}(-{\bf G}^{\tau,\tau''}_{-J+1,-J}+{\bf G}^{\tau,\tau''}_{J-1,J}) \nonumber \\
	&+&...+
	\sqrt{3} (-{\bf G}^{\tau,\tau''}_{J-1,J-2}+{\bf G}^{\tau,\tau''}_{-J+1,-J+2})+
	(-{\bf G}^{\tau,\tau''}_{J,J-1}+{\bf G}^{\tau,\tau''}_{-J,-J+1})\Big),
	\label{ap_Hw2y} \\
	\mathrm{i}{\bf H}_{\omega_2,z}&=&\mu_a\Big(\sqrt{2J+1}({\bf G}^{\tau,\tau''}_{-J,-J}+{\bf G}^{\tau,\tau''}_{J,J})+\sqrt{4J}({\bf G}^{\tau,\tau''}_{-J+1,-J+1}+{\bf G}^{\tau,\tau''}_{J-1,J-1}) \nonumber \\
	&+&...+
	(J+1){\bf G}^{\tau,\tau''}_{0,0}\Big).
	\label{ap_Hw2z}
	\end{eqnarray}
\end{widetext}
Labeling the three rotational levels by $\tau$, $\tau'$ and $\tau''$,
  we denote the generalized Pauli matrices that describe the interaction between the states $|J,\tau,M\rangle$ and $|J+1,\tau',M' \rangle$ as ${\bf G}^{\tau,\tau'}_{M,M'}$ and
	${\bf F}^{\tau,\tau'}_{M,M'}$, and the interaction between the states $|J,\tau,M\rangle$ and $|J+1,\tau'',M' \rangle$ as ${\bf G}^{\tau,\tau''}_{M,M'}$ and
	${\bf F}^{\tau,\tau''}_{M,M'}$. These matrices correspond to ${\bf G}_{jk}$ and ${\bf F}_{jk}$ in Eq.~\eqref{basis}.
	The interaction Hamiltonians \eqref{ap_Hw1x}, \eqref{ap_Hw1y}, and \eqref{ap_Hw2y} are linear combinations of $(2J+1)$-pairs of generalized Pauli matrices with different prefactors, while Eq.~\eqref{ap_Hw2z} is a sum of $J$-pairs plus a single element.
In order to carry out the proof, we adapt the graph representation introduced in Ref.~\cite{BCCS,BCS} to the asymmetric quantum rotor. Graph representations have also been used to study the controllability of quantum walks~\cite{GodsilPRA10} and quantum networks~\cite{GoklerPRL17} in quantum information. 
In the present case of the asymmetric quantum rotor, the graph representation together with a Lie algebraic tool based on the properties of the Vandermonde matrix (see Eq.(\ref{ap_Vandermonde})) is crucial to isolate the Lie algebra basis elements and thus find the basis for the induction.
The graph is obtained by presenting 
the eigenvalues of ${\bf H}_0$ as vertices (indicated by the horizontal bars in Fig.~\ref{scheme_proof_general}). The edges of the graph (colored lines in Fig.~\ref{scheme_proof_general}) are given by the generalized Pauli matrices that occur in ${\bf H}_{\omega_i,p}$, cf. Eqs.~\eqref{ap_Hw1x} - \eqref{ap_Hw2z}. Note that lines with same color belong to the same interaction Hamiltonian ${\bf H}_{\omega_i,p}$ The graph shown in panel (a) of Fig.~\ref{scheme_proof_general} presents ${\bf H}_0$ interacting with the control Hamiltonians 
${\bf H}_{\omega_1,x}$ or ${\bf H}_{\omega_1,y}$. The two Hamiltonians describe the same transitions and only differ by the relative signs of the transitions, such that adding and subtracting the two Hamiltonians leads to the distinct graphs depicted by the red and blue lines. Panels (b) and (c) present the graphs for the interaction with ${\bf H}_{\omega_2,z}$ and
${\bf H}_{\omega_2,y}$, respectively.

 \section{Generating the Lie algebra of an arbitrary rotational subsystem by induction}
 \label{sec:induction}

To prove  Eq.~\eqref{su(n)}, we repeatedly take commutators and linear combinations of Eqs.~\eqref{ap_Hw1x}-\eqref{ap_Hw2z} and $\mathrm{i}{\bf H}_{0}$, to show that each of the generalized Pauli matrices, or basis elements, that occurs in Eq.~\eqref{ap_Hw1x}-\eqref{ap_Hw2z} is an element of $\mathrm{L}$.
Realizing that the basis elements in Eqs.~\eqref{ap_Hw1x})-\eqref{ap_Hw2z} create a connected graph, we can conclude that the remaining basis elements of $\mathfrak{su}(6J+7)$ are also in $\mathrm{L}$. 
The proof is divided into six steps.

\subsection*{Step 1: Isolating the basis elements occurring in $\mathrm{i}{\bf H}_{\omega_1,x}$ and
  $\mathrm{i}{\bf H}_{\omega_1,y}$}

We construct Hamiltonians $\mathrm{i} {\bf H}_{\omega_1,\sigma_\pm}$
as linear combinations of  operators which are in $\mathrm{L}$,
\begin{widetext}
	\begin{eqnarray}
	\mathrm{i} {\bf H}_{\omega_1,\sigma_+} &:=&\frac{1}{2}\left(\mathrm{i}{\bf H }_{\omega_1,x}+ [\mathrm{i}{\bf H}_{0},\mathrm{i}{\bf H}_{\omega_1,y}]/\omega_1\right) \nonumber \\
	&=& \mu_c\Big(\sqrt{(J+1)(2J+1)}{\bf G}^{\tau,\tau'}_{J,J+1}+\sqrt{J(2J+1)}{\bf G}^{\tau,\tau'}_{J-1,J}+...+\sqrt{3} {\bf G}^{\tau,\tau'}_{-J+1,-J+2} +
	{\bf G}^{\tau,\tau'}_{-J,-J+1}\Big)
	\label{ap_sigp}
	\end{eqnarray}
	and
	\begin{eqnarray}
	\mathrm{i} {\bf H}_{\omega_1,\sigma_-} &:=& \frac{1}{2}\left(\mathrm{i}{\bf H }_{\omega_1,x}- [\mathrm{i}{\bf H}_{0},\mathrm{i}{\bf H}_{\omega_1,y}]/\omega_1\right) \nonumber \\
	&=& \mu_c\Big(\sqrt{(J+1)(2J+1)}{\bf G}^{\tau,\tau'}_{-J,-(J+1)}+\sqrt{J(2J+1)}{\bf G}^{\tau,\tau'}_{-J+1,-J}+...+ \sqrt{3} {\bf G}^{\tau,\tau'}_{J-1,J-2} +
	{\bf G}^{\tau,\tau'}_{J,J-1}\Big)\,,
	\label{ap_sigm}
	\end{eqnarray}	
\end{widetext}
where we have used Eq.~\eqref{relation4} to compute the commutators.
The Hamiltonians $\mathrm{i} {\bf H}_{\omega_1,\sigma_\pm}$ describe the interaction with right and left circularly polarized radiation with frequency $\omega_1$, and 
the operators in Eqs.~\eqref{ap_sigp} and \eqref{ap_sigm} contain only those generalized Pauli matrices which correspond to the blue and red lines in Fig.~\ref{scheme_proof_general} (a). 
Using the abbreviations $\mathrm{ad}^{n+1}_A B=[A,\mathrm{ad}^{n}_A B]$ and $\mathrm{ad}^{0}_A B=B$ and 
defining ${\bf J}(\mathrm{i}{\bf H}_{\omega_1,\sigma_+})=[\mathrm{i}{\bf H}_{0} ,\mathrm{i}{\bf H}_{\omega_1,\sigma_+}]/\omega_1$, we find 
\begin{widetext}
  \begin{eqnarray*}
    \mathrm{ad}^{2s}_{\bf J({\rm i}{\bf H}_{\omega_1,\sigma_+})} {\rm i}{\bf H}_{\omega_1,\sigma_+} 
    \propto \Big(\sqrt{(J+1)(2J+1)}^{2s+1}{\bf G}^{\tau,\tau'}_{J,J+1}+\sqrt{J(2J+1)}^{2s+1}{\bf G}^{\tau,\tau'}_{J-1,J}+...+
    \sqrt{3}^{2s+1} {\bf G}^{\tau,\tau'}_{-J+1,-J+2}
    +{\bf G}^{\tau,\tau'}_{-J,-J+1}\Big)
  \end{eqnarray*}
  for $s=0,\dots,2J$.
	We can thus write 
	\begin{eqnarray}
	\begin{pmatrix}
	\mathrm{ad}^{0}_{\bf J({\rm i}{\bf H}_{\omega_1,\sigma_+})} {\rm i}{\bf H}_{\omega_1,\sigma_+} \\
	\mathrm{ad}^{2}_{\bf J({\rm i}{\bf H}_{\omega_1,\sigma_+})} {\rm i}{\bf H}_{\omega_1,\sigma_+} \\
	\vdots \\
	\mathrm{ad}^{4J-2}_{\bf J({\rm i}{\bf H}_{\omega_1,\sigma_+})} {\rm i}{\bf H}_{\omega_1,\sigma_+} \\
	\mathrm{ad}^{4J}_{\bf J({\rm i}{\bf H}_{\omega_1,\sigma_+})} {\rm i}{\bf H}_{\omega_1,\sigma_+}
	\end{pmatrix}=V \begin{pmatrix}
	{\bf G}^{\tau,\tau'}_{J,J+1}\\
	{\bf G}^{\tau,\tau'}_{J-1,J}\\
	\vdots\\
	{\bf G}^{\tau,\tau'}_{-J+1,-J+2}\\
	{\bf G}^{\tau,\tau'}_{-J,-J+1} 
	\end{pmatrix}
	\label{ap_Vandermonde}
	\end{eqnarray}
	with
	$$
	\qquad V=\begin{pmatrix}
	\sqrt{(J+1)(2J+1)} & \sqrt{J(2J+1)} & \cdots &\sqrt{3}& 1\\
	\sqrt{(J+1)(2J+1)}^{3} & \sqrt{J(2J+1)}^{3} & \cdots &\sqrt{3}^{3}& 1\\
	\sqrt{(J+1)(2J+1)}^{5} & \sqrt{J(2J+1)}^{5} &  &\sqrt{3}^{5}& 1\\
	\vdots &  &  & & \\
	\sqrt{(J+1)(2J+1)}^{4J-1} & \sqrt{J(2J+1)}^{4J-1} & &\sqrt{3}^{4J-1}& 1 \\
	\sqrt{(J+1)(2J+1)}^{4J+1} & \sqrt{J(2J+1)}^{4J+1} & &\sqrt{3}^{4J+1}& 1
	\end{pmatrix}. $$
\end{widetext}
Since $V$ is a Vandermonde matrix, its determinant is given by the product of the sum and the difference of every pair of the coefficients in the first row.
Noticing that those coefficients form a positive, strictly increasing sequence, we see that they are all different. Thus $V$ is invertible, and we find that
\begin{eqnarray}\label{elements1}
{\bf G}^{\tau,\tau'}_{J,J+1}, {\bf G}^{\tau,\tau'}_{J-1,J}, 
...,
{\bf G}^{\tau,\tau'}_{-J+1,-J+2},
{\bf G}^{\tau,\tau'}_{-J,-J+1} 
\in \mathrm{L}\,, \nonumber \\
\end{eqnarray}
Replacing  
${\rm i}{\bf H}_{\omega_1,\sigma_+}$ by ${\rm i}{\bf H}_{\omega_1,\sigma_-}$ in Eq.~\eqref{ap_Vandermonde}, we find analogously that
\begin{eqnarray}\label{elements2}
{\bf G}^{\tau,\tau'}_{-J,-(J+1)},{\bf G}^{\tau,\tau'}_{-J+1,-J},...,
{\bf G}^{\tau,\tau'}_{J-1,J-2},
{\bf G}^{\tau,\tau'}_{J,J-1} \in \mathrm{L}. \nonumber \\
\end{eqnarray}
We have thus shown that each of the basis elements indicated by the blue and red lines in
Fig.~\ref{scheme_proof_general} is an element of $\mathrm{L}$.

\subsection*{Step 2: Isolating the basis elements occurring in $\mathrm{i}{\bf H}_{\omega_2,z}$}
  
\begin{figure}[tbp]
  \includegraphics[width=1.0\linewidth]{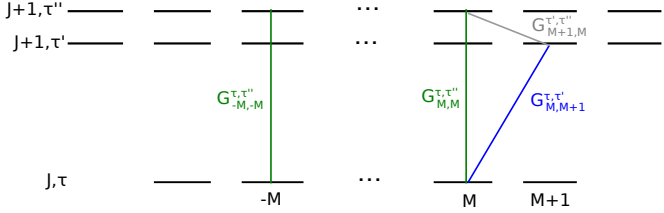} 
  \caption{Illustration of the double commutator Eq.~\eqref{elements4}: The commutator between ${\bf G}^{\tau,\tau''}_{-M,-M}+{\bf G}^{\tau,\tau''}_{M,M}$ (green lines) and ${\bf G}^{\tau,\tau'}_{M,M+1}$ (blue line) results in the basis element indicated by the gray line. The commutator between the basis elements represented by the gray and blue lines then results in ${\bf G}^{\tau,\tau''}_{M,M}$ (right green line) alone.}
  \label{commutator_1}
\end{figure}
We now reproduce the previous argument for the operator ${\rm i}{\bf H}_{\omega_2,z}$. Replacing  
${\rm i}{\bf H}_{\omega_1,\sigma_+}$ by ${\rm i}{\bf H}_{\omega_2,z}$ in Eq.~\eqref{ap_Vandermonde}, and noticing that in this case the sequence of coefficients in the first row of the corresponding matrix $V$ is positive and strictly decreasing, we find that
\begin{eqnarray}\label{elements3}
{\bf G}^{\tau,\tau''}_{-J,-J}+{\bf G}^{\tau,\tau''}_{J,J},
{\bf G}^{\tau,\tau''}_{-J+1,-J+1}+
{\bf G}^{\tau,\tau''}_{J-1,J-1}, \nonumber \\
...,{\bf G}^{\tau,\tau''}_{0,0} \in \mathrm{L}.
\end{eqnarray}
To separate the sum over $M$ from that over $-M$ in \eqref{elements3}, we take double commutators with matrices the of Eq.~\eqref{elements1}, that is,
\begin{equation}
\left[\left[{\bf G}^{\tau,\tau''}_{-M,-M}+{\bf G}^{\tau,\tau''}_{M,M}, {\bf G}^{\tau,\tau'}_{M,M+1} \right], {\bf G}^{\tau,\tau'}_{M,M+1}\right] =-{\bf G}^{\tau,\tau''}_{M,M} \label{double_commutator}\,,
\end{equation}
which is also illustrated in Fig.~\ref{commutator_1}.
Thus 
\begin{eqnarray}\label{elements4}
{\bf G}^{\tau,\tau''}_{-J,-J},...,{\bf G}^{\tau,\tau''}_{J,J} \,,
\in \mathrm{L},
\end{eqnarray}
i.e., all basis elements indicated by the green lines in Fig.~\ref{scheme_proof_general}(b)
are elements of $\mathrm{L}$.
Note, that instead of calculating the double commutators as in  Eq.~\eqref{double_commutator}, one could also graphically deduce the basis elements: The double commutator between a linear combination of basis elements (indicated by the green lines in Fig.~\ref{commutator_1}), and a single basis element (indicated by the blue line) contains only those basis elements of the linear combination, which have a common vertex with the single basis element. We will extensively use this technique in the following steps of the proof.

\subsection*{Step 3:  Isolating the basis elements occurring in $\mathrm{i}{\bf H}_{\omega_2,y}$}

\begin{figure*}[tb]
	\includegraphics[width=14cm]{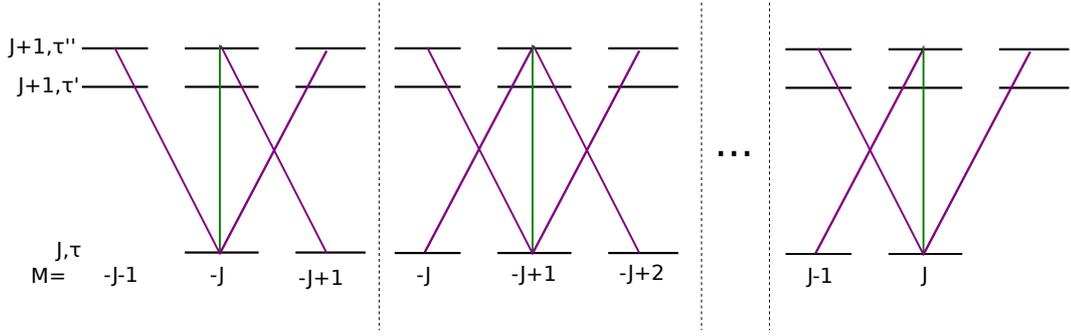} 
	\caption{Illustration of the double commutator between $\mathrm{i}{\bf H}_{\omega_2,y}$ and the basis elements (\ref{elements4}), depicted as green lines: The double commutator between $\mathrm{i}{\bf H}_{\omega_2,y}$ and ${\bf G}^{\tau,\tau''}_{-J,-J}$ results in an operator, which contains the three purple lines shown in the left panel. The four purple lines in the next panel depict the result of the double commutator between $\mathrm{i}{\bf H}_{\omega_2,y}$ and ${\bf G}^{\tau,\tau''}_{-J+1,-J+1}$, and so on.}
	\label{commutator_2}
\end{figure*}
Next, we isolate the basis elements that occur in interaction Hamiltonian $\mathrm{i}{\bf H}_{\omega_2,y}$, i.e., the purple lines in Fig.~\ref{scheme_proof_general}(c), by means of a graph proof.
Taking double commutators of $\mathrm{i}{\bf H}_{\omega_2,y}$ with the basis elements obtained in Eq.~\eqref{elements3}, we can isolate $2J+1$ groups of interactions within ${\rm i}{\bf H}_{\omega_2,y}$, where each group is centered around the transition 
$$(J,\tau,M)\leftrightarrow (J+1,\tau'',M), \quad M=-J,\dots,J. $$
This is illustrated in Fig.~\eqref{commutator_2}.
We find for all $M\neq \pm J$, 
\begin{widetext}
	\begin{equation}\label{M}
	\begin{split}
	\left[\left[{\rm i}{\bf H}_{\omega_2,y},{\bf G}^{\tau,\tau''}_{M,M}\right],{\bf G}^{\tau,\tau''}_{M,M}\right] = &-\sqrt{\dfrac{1}{2}(J+M+1)(J+M)}{\bf G}^{\tau,\tau''}_{M-1,M}-\sqrt{\dfrac{1}{2}(J+M+2)(J+M+1)}{\bf G}^{\tau,\tau''}_{M,M+1}\\ &
	+\sqrt{\dfrac{1}{2}(J-M+1)(J-M)}{\bf G}^{\tau,\tau''}_{M+1,M}+\sqrt{\dfrac{1}{2}(J-M+2)(J-M+1)}{\bf G}^{\tau,\tau''}_{M,M-1}\,,
	\end{split}
	\end{equation}
	with the resulting four generalized Pauli matrices indicated by the purple lines in the second panel from the left in Fig.~\ref{commutator_2}.
	If $M=-J$,
	\begin{equation}\label{-J}
	\left[\left[{\rm i}{\bf H}_{\omega_2,y},{\bf G}^{\tau,\tau''}_{-J,-J}\right],{\bf G}^{\tau,\tau''}_{-J,-J}\right] =\sqrt{(J+1)(2J+1)}{\bf G}^{\tau,\tau''}_{-J,-J-1}+\sqrt{J(2J+1)}{\bf G}^{\tau,\tau''}_{-J+1,-J}-{\bf G}^{\tau,\tau''}_{-J,-J+1}\,,
	\end{equation}
where three generalized Pauli matrices are shown as purple lines in the left panel of Fig.~\ref{commutator_2}. Finally, if $M=J$,
	\begin{equation}\label{J}
	\left[\left[{\rm i}{\bf H}_{\omega_2,y},{\bf G}^{\tau,\tau''}_{J,J}\right],{\bf G}^{\tau,\tau''}_{J,J}\right] =-\sqrt{(J+1)(2J+1)}{\bf G}^{\tau,\tau''}_{J,J+1}-\sqrt{J(2J+1)}{\bf G}^{\tau,\tau''}_{J-1,J}+{\bf G}^{\tau,\tau''}_{J,J-1}\,,
	\end{equation}
\end{widetext}
with  the three generalized Pauli matrices shown in the right panel of Fig.~\ref{commutator_2}. 

Next, we show by induction on $M$ that each of the purple lines in Fig.~\ref{commutator_2} can be isolated. As basis for the inductive argument, we first show that the transitions around 
$(J,\tau,-J)\leftrightarrow (J+1,\tau'',-J)$ and $(J,\tau,-J+1)\leftrightarrow (J+1,\tau'',-J+1)$, indicated by the purple lines in the left and second-left panel of Fig.~\ref{commutator_2}, can be isolated. We then carry out the inductive step, that is, we prove that, if we can isolate each of the four basis elements around the transition $(J,\tau,M)\leftrightarrow (J+1,\tau'',M)$, then we can do the same for the basis elements around the transition $(J,\tau,M+1)\leftrightarrow (J,\tau'',M+1)$ for all $M<J-1$.

\subsection*{Step 4: Basis of induction}

\begin{figure*}[tbp]
	\includegraphics[width=1.0\linewidth]{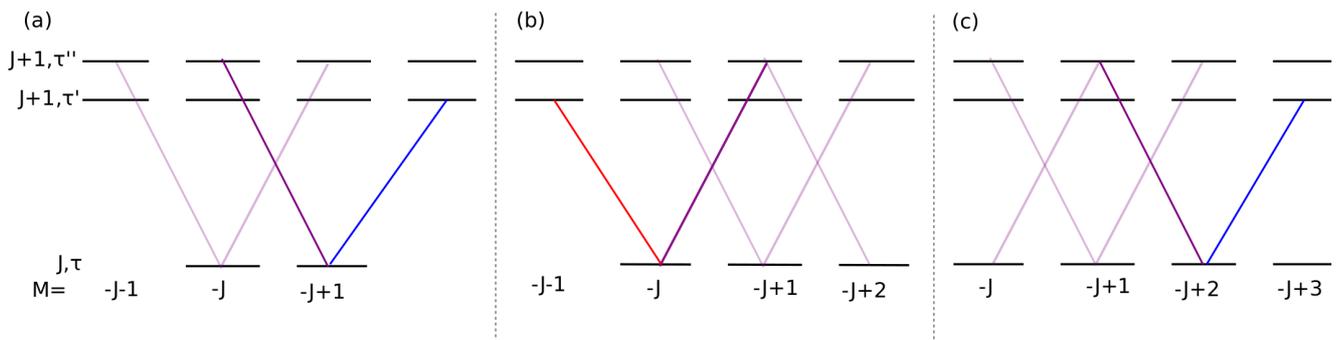} 
	\caption{A linear combination of the basis elements depicted by the (light and dark) purple lines is an operator $\in \mathrm{L}$. The basis elements depicted by the dark purple lines in (a), (b), and (c) can be isolated by calculating the double commutator with the basis element depicted by the blue (a, c) and red (b) lines.}
	\label{commutator_3}
\end{figure*}
Since ${\bf G}^{\tau,\tau'}_{-J+1,-J+2} \in \mathrm{L}$, cf. Eq.~\eqref{elements1}, we start by computing the double commutator of \eqref{-J} with ${\bf G}^{\tau,\tau'}_{-J+1,-J+2}$.
As indicated in Fig.~\ref{commutator_3}(a), this operation yields 
\begin{equation}\label{one}
{\bf G}^{\tau,\tau''}_{-J+1,-J}\in \mathrm{L}\,.
\end{equation}
Moreover, according to Eq.~\eqref{elements2}, we can compute the double commutators of \eqref{M} for $M=-J+1$ with ${\bf G}^{\tau,\tau'}_{-J,-J-1}$. The action of this double commutator is depicted in Fig.~\ref{commutator_3}(b) and results in 
\begin{equation}\label{two}
{\bf G}^{\tau,\tau''}_{-J,-J+1}\in \mathrm{L}\,.
\end{equation}
Taking the double commutator of \eqref{M} for $M=-J+1$ with ${\bf G}^{\tau,\tau'}_{-J+2,-J+3}$ we find that
\begin{equation}\label{three}
{\bf G}^{\tau,\tau''}_{-J+2,-J+1}\in \mathrm{L}\,,
\end{equation}
which is illustrated in  Fig.~\ref{commutator_3}(c).
Now, subtracting a suitable linear combination of Eqs.~\eqref{one}, \eqref{two}, and \eqref{three} from \eqref{M} for $M=-J+1$ results in 
\begin{equation}\label{four}
{\bf G}^{\tau,\tau''}_{-J+1,-J+2}\in \mathrm{L}\,.
\end{equation}
We have thus shown that the generalized Pauli matrices corresponding to the four purple lines in the second-left panel of Fig.~\ref{commutator_2} can be isolated.
Subtracting a suitable linear combination of Eqs.~\eqref{one} and \eqref{two} from \eqref{-J}, we find that
\begin{equation}\label{five}
{\bf G}^{\tau,\tau''}_{-J,-J-1}\in \mathrm{L}\,.
\end{equation}
Thus, also the three generalized Pauli matrices indicated by the purple lines in the first panel of  Fig.~\ref{commutator_2} can be isolated.
This concludes the basis of the induction.

\subsection*{Step 5: Inductive step}

\begin{figure}[tbp]
  \includegraphics[width=1.0\linewidth]{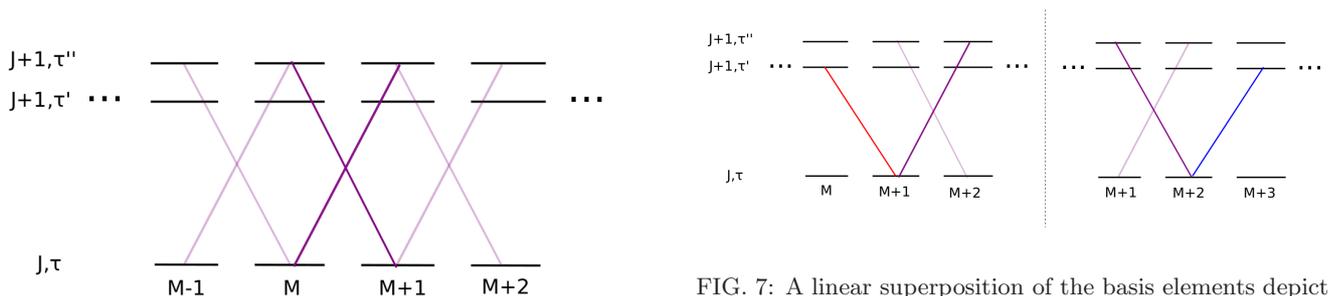} 
  \caption{The dark purple lines are part of the set of basis elements centered around the $(J,\tau,M) \leftrightarrow (J+1,\tau'',M)$-transition as well as of the set of basis elements centered around the $(J,\tau,M+1) \leftrightarrow (J+1,\tau'',M+1)$-transition.}
  \label{commutator_4}
\end{figure}
\begin{figure}[tbp]
	\includegraphics[width=1.0\linewidth]{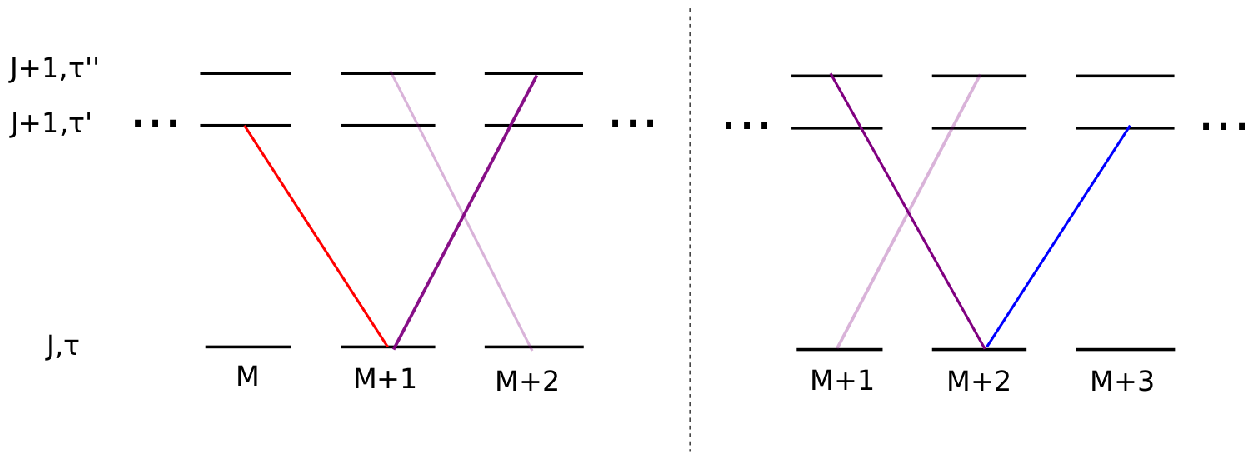} 
	\caption{A linear superposition of the basis elements depicted by the (light and dark) purple lines is an operator $\in \mathrm{L}$. The basis elements depicted by the dark purple lines in both panels can be isolated by calculating the double commutator with the basis element depicted by the blue and red lines.}
	\label{commutator_5}
\end{figure}

We now prove the inductive step, that is, if we can isolate each of the basis elements presented by the four lines around the transition
$(J,\tau,M)\leftrightarrow (J+1,\tau'',M)$, then we can do the same with the basis elements around the transition $(J,\tau,M+1)\leftrightarrow (J+1,\tau'',M+1)$ for all $M<J-1$. Indeed, inspection of 
Fig.~\ref{commutator_4} reveals that the transitions
$(J,\tau,M)\leftrightarrow (J+1,\tau'',M+1)$ and $(J,\tau,M+1)\leftrightarrow (J+1,\tau'',M)$ are  common for both sets of transitions. Thus the inductive hypothesis implies that we are left to show that  the sum of basis elements
\[
\begin{split}
&\sqrt{\dfrac{1}{2}(J+M+3)(J+M+2)}{\bf G}^{\tau,\tau''}_{M+1,M+2}\\ &
+\sqrt{\dfrac{1}{2}(J-M)(J-M-1)}{\bf G}^{\tau,\tau''}_{M+2,M+1} \in \mathrm{L}
\end{split}
\]
can be separated.
This can be done by taking double commutators with ${\bf G}^{\tau,\tau'}_{M+2,M+3} \in \mathrm{L}$ and ${\bf G}^{\tau,\tau'}_{M+1,M} \in \mathrm{L}$, as illustrated in Fig. \ref{commutator_5}. Thus, it remains to be shown that the basis elements depicted by purple lines in the right panel in Fig.~\ref{commutator_3} can be isolated. Since it has already been shown that the  basis elements corresponding to the transitions
$(J,\tau,J-1) \leftrightarrow (J+1,\tau'',J)$ and  $(J,\tau,J) \leftrightarrow (J+1,\tau'',J-1)$ can be isolated, the remaining basis element corresponding to the transition $(J,\tau,J) \leftrightarrow (J+1,\tau'',J+1)$ can be isolated by subtracting these two elements. 
We have thus demonstrated that all generalized Pauli matrices appearing in ${\rm i}{\bf H}_{\omega_2,y}$, i.e. all basis elements depicted by purple lines in Fig.~\ref{scheme_proof_general}(c) are in $\mathrm{L}$.

\subsection*{Step 6: Connectedness}

In the previous steps, we have shown that each basis element present in Eqs.~\eqref{ap_Hw1x}--\eqref{ap_Hw2z} belongs to $\mathrm{L}$. We are left to prove that the remaining Pauli matrices spanning $\mathfrak{su}(6J+7)$ are in $\mathrm{L}$ as well.
As one can see from Fig.~\ref{scheme_proof_general}, the lines 
(or \emph{edges}, in graph theoretical terminology) representing the basis elements present in Eqs.~\eqref{ap_Hw1x}--\eqref{ap_Hw2z},
form a connected graph. In other words, any pair of rotational eigenstates can be connected by following blue, red, and purple lines. 
It follows from
Eqs.~\eqref{eq:commutators}
that, given two concatenated edges of the graph, 
the commutator between their corresponding basis elements is another basis element. 
The edge associated with this new basis element connects the external vertices of the two concatenated edges. 
The new basis element is also in $L$, since the latter is a Lie algebra.

Iterating this reasoning for longer and longer concatenations of edges, 
we find that $L$  contains all generalized Pauli matrices ${\bf G}^{\sigma,\nu}_{M,M'}$, for $\sigma,\nu=\tau,\tau',\tau''$ and $-J \leq M,M' \leq J$ if $\sigma$ or $\nu$ is equal to $\tau$ and $-(J+1) \leq M,M' \leq (J+1)$ otherwise.
By applying relations~\eqref{eq:commutators}, we find that all  matrices  ${\bf F}_{M,M'}^{\sigma,\nu}$ and ${\bf D}_{M,M'}^{\sigma,\nu}$ are in $L$ as well. This concludes the proof of Eq.~\eqref{su(n)}. 

\section{conclusions}\label{sec:concl}

We have presented a method to construct the basis of the Lie algebra for a highly degenerate, three-level rotational subsystem $J/J+1/J+1$ of an asymmetric top with arbitrarily high rotational excitation. This is a prerequisite of proving controllability of such subsystems.
The controllability of the complete spectrum of an asymmetric top has been analyzed in a perturbative treatment~\cite{Pozzoli2021}. Controlling a particular subsystem of an asymmetric top  is often both necessary and sufficient in view of applications~\cite{Leibscher20}. In practice, the subsystem can typically be isolated from the rest of the Hilbert space by fulfilling the corresponding spectral condition. In case of an asymmetric top, this is realized by choosing frequencies and intensities of the (microwave) radiation such that only few rotational transitions are addressed \cite{EibenbergerPRL17,PerezAngewandte17,PerezJPCL18}.
We have generalized here the result of Ref.~\cite{Leibscher20} 
on the minimal number of fields for the rotational subsystem to be controllable from $J=0$ and $J=1$ to arbitrary $J$. This was made possible by making use of a graph representation similar to that in Refs.~\cite{BCS, GodsilPRA10}.
Presenting the eigenvalues of the system as edges and the transitions induced by the control fields as vertices of a graph has allowed us to determine all nested commutators via an inductive argument and thus construct the basis of the rotational subsystem's Lie algebra for arbitrary $J$. This is a necessary prerequisite to analyze controllability of arbitrary rotational subsystems \cite{Leibscher20}. Analyzing the controllability of the rotational subsystems considered here is of practical importance for current applications of quantum asymmetric top rotors from quantum information~\cite{AlbertPRX20}
to high-resolution spectroscopy~\cite{DomingosAnnuRevPhysChem18}. 

Our approach combining a graphical representation of the Hamiltonian with an inductive construction of the dynamical Lie algebra can in principle be applied to other Hamiltonians defined on a Hilbert space with tensor sum structure. 
Furthermore, we believe that extension to Hamiltonians defined on a tensor Hilbert space may also be possible. In this case, the treatment of interactions represents a challenge, in addition to a potentially large Hilbert space with many degenerate levels. Overcoming this challenge would allow us to
advance present understanding of controllability of arrays of interacting two-level systems~\cite{SchirmerPRA08,WangIEEE12,WangPRA16,GoklerPRL17,ChenPRA20,Albertini21} by, for example, identifying the drives that are needed to implement any unitary evolution in such arrays. 
This is the subject of future work.

\begin{acknowledgments}
	We gratefully acknowledge financial support from the Deutsche Forschungsgemeinschaft through CRC 1319 ELCH and from the
	European Union's Horizon 2020 research and innovation programme
	under the Marie Sklodowska-Curie grant agreement Nr. 765267 (QuSCo). 
	MS and UB also thank the ANR projects SRGI ANR-15-CE40-0018 and Quaco ANR-17-CE40-0007-01. 
\end{acknowledgments}


\begin{thebibliography}{29}
	\expandafter\ifx\csname natexlab\endcsname\relax\def\natexlab#1{#1}\fi
	\expandafter\ifx\csname bibnamefont\endcsname\relax
	\def\bibnamefont#1{#1}\fi
	\expandafter\ifx\csname bibfnamefont\endcsname\relax
	\def\bibfnamefont#1{#1}\fi
	\expandafter\ifx\csname citenamefont\endcsname\relax
	\def\citenamefont#1{#1}\fi
	\expandafter\ifx\csname url\endcsname\relax
	\def\url#1{\texttt{#1}}\fi
	\expandafter\ifx\csname urlprefix\endcsname\relax\def\urlprefix{URL }\fi
	\providecommand{\bibinfo}[2]{#2}
	\providecommand{\eprint}[2][]{\url{#2}}
	
	\bibitem[{\citenamefont{Gilmore}(2008)}]{GilmoreBook}
	\bibinfo{author}{\bibfnamefont{R.}~\bibnamefont{Gilmore}},
	\emph{\bibinfo{title}{Lie Groups, Physics, and Geometry}}
	(\bibinfo{publisher}{Cambridge University Press}, \bibinfo{year}{2008}).
	
	\bibitem[{\citenamefont{D'Alessandro}(2008)}]{Alessandro2008}
	\bibinfo{author}{\bibfnamefont{D.}~\bibnamefont{D'Alessandro}},
	\emph{\bibinfo{title}{Quantum Control and Dynamics}}
	(\bibinfo{publisher}{Chapman and Hall (CRC)}, \bibinfo{year}{2008}).
	
	\bibitem[{\citenamefont{Glaser et~al.}(2015)\citenamefont{Glaser, Boscain,
			Calarco, Koch, K\"ockenberger, Kosloff, Kuprov, Luy, Schirmer,
			Schulte-Herbr\"uggen et~al.}}]{GlaserEPJD15}
	\bibinfo{author}{\bibfnamefont{S.~J.} \bibnamefont{Glaser}},
	\bibinfo{author}{\bibfnamefont{U.}~\bibnamefont{Boscain}},
	\bibinfo{author}{\bibfnamefont{T.}~\bibnamefont{Calarco}},
	\bibinfo{author}{\bibfnamefont{C.~P.} \bibnamefont{Koch}},
	\bibinfo{author}{\bibfnamefont{W.}~\bibnamefont{K\"ockenberger}},
	\bibinfo{author}{\bibfnamefont{R.}~\bibnamefont{Kosloff}},
	\bibinfo{author}{\bibfnamefont{I.}~\bibnamefont{Kuprov}},
	\bibinfo{author}{\bibfnamefont{B.}~\bibnamefont{Luy}},
	\bibinfo{author}{\bibfnamefont{S.}~\bibnamefont{Schirmer}},
	\bibinfo{author}{\bibfnamefont{T.}~\bibnamefont{Schulte-Herbr\"uggen}},
	\bibnamefont{et~al.}, \bibinfo{journal}{Eur. Phys. J. D}
	\textbf{\bibinfo{volume}{69}}, \bibinfo{pages}{279} (\bibinfo{year}{2015}).
	
	\bibitem[{\citenamefont{Fu et~al.}(2001)\citenamefont{Fu, Schirmer, and
			Solomon}}]{Schirmer}
	\bibinfo{author}{\bibfnamefont{H.}~\bibnamefont{Fu}},
	\bibinfo{author}{\bibfnamefont{S.~G.} \bibnamefont{Schirmer}},
	\bibnamefont{and} \bibinfo{author}{\bibfnamefont{A.~I.}
		\bibnamefont{Solomon}}, \bibinfo{journal}{J. Phys. A}
	\textbf{\bibinfo{volume}{34}}, \bibinfo{pages}{1679} (\bibinfo{year}{2001}).
	
	\bibitem[{\citenamefont{Albert et~al.}(2020)\citenamefont{Albert, Covey, and
			Preskill}}]{AlbertPRX20}
	\bibinfo{author}{\bibfnamefont{V.~V.} \bibnamefont{Albert}},
	\bibinfo{author}{\bibfnamefont{J.~P.} \bibnamefont{Covey}}, \bibnamefont{and}
	\bibinfo{author}{\bibfnamefont{J.}~\bibnamefont{Preskill}},
	\bibinfo{journal}{Phys. Rev. X} \textbf{\bibinfo{volume}{10}},
	\bibinfo{pages}{031050} (\bibinfo{year}{2020}).
	
	\bibitem[{\citenamefont{Domingos et~al.}(2018)\citenamefont{Domingos,
			P{\'e}rez, and Schnell}}]{DomingosAnnuRevPhysChem18}
	\bibinfo{author}{\bibfnamefont{S.~R.} \bibnamefont{Domingos}},
	\bibinfo{author}{\bibfnamefont{C.}~\bibnamefont{P{\'e}rez}},
	\bibnamefont{and} \bibinfo{author}{\bibfnamefont{M.}~\bibnamefont{Schnell}},
	\bibinfo{journal}{Annu. Rev. Phys. Chem.} \textbf{\bibinfo{volume}{69}},
	\bibinfo{pages}{499} (\bibinfo{year}{2018}).
	
	\bibitem[{\citenamefont{Brumer and Shapiro}(2003)}]{ShapiroBook}
	\bibinfo{author}{\bibfnamefont{P.}~\bibnamefont{Brumer}} \bibnamefont{and}
	\bibinfo{author}{\bibfnamefont{M.}~\bibnamefont{Shapiro}},
	\emph{\bibinfo{title}{Principles and Applications of the Quantum Control of
			Molecular Processes}} (\bibinfo{publisher}{Wiley Interscience},
	\bibinfo{year}{2003}).
	
	\bibitem[{\citenamefont{Judson et~al.}(1990)\citenamefont{Judson, Lehmann,
			Rabitz, and Warren}}]{Judson1990}
	\bibinfo{author}{\bibfnamefont{R.}~\bibnamefont{Judson}},
	\bibinfo{author}{\bibfnamefont{K.}~\bibnamefont{Lehmann}},
	\bibinfo{author}{\bibfnamefont{H.}~\bibnamefont{Rabitz}}, \bibnamefont{and}
	\bibinfo{author}{\bibfnamefont{W.}~\bibnamefont{Warren}},
	\bibinfo{journal}{Journal of Molecular Structure}
	\textbf{\bibinfo{volume}{223}}, \bibinfo{pages}{425} (\bibinfo{year}{1990}).
	
	\bibitem[{\citenamefont{Chambrion}(2012)}]{chambrion}
	\bibinfo{author}{\bibfnamefont{T.}~\bibnamefont{Chambrion}},
	\bibinfo{journal}{Automatica J. IFAC} \textbf{\bibinfo{volume}{48}},
	\bibinfo{pages}{2040} (\bibinfo{year}{2012}).
	
	\bibitem[{\citenamefont{Chambrion et~al.}(2009)\citenamefont{Chambrion, Mason,
			Sigalotti, and Boscain}}]{CMSB}
	\bibinfo{author}{\bibfnamefont{T.}~\bibnamefont{Chambrion}},
	\bibinfo{author}{\bibfnamefont{P.}~\bibnamefont{Mason}},
	\bibinfo{author}{\bibfnamefont{M.}~\bibnamefont{Sigalotti}},
	\bibnamefont{and} \bibinfo{author}{\bibfnamefont{U.}~\bibnamefont{Boscain}},
	\bibinfo{journal}{Ann. Inst. H. Poincar\'{e} Anal. Non Lin\'{e}aire}
	\textbf{\bibinfo{volume}{26}}, \bibinfo{pages}{329} (\bibinfo{year}{2009}).
	
	\bibitem[{\citenamefont{Boussa\"{\i}d et~al.}(2013)\citenamefont{Boussa\"{\i}d,
			Caponigro, and Chambrion}}]{MR3101605}
	\bibinfo{author}{\bibfnamefont{N.}~\bibnamefont{Boussa\"{\i}d}},
	\bibinfo{author}{\bibfnamefont{M.}~\bibnamefont{Caponigro}},
	\bibnamefont{and}
	\bibinfo{author}{\bibfnamefont{T.}~\bibnamefont{Chambrion}},
	\bibinfo{journal}{IEEE Trans. Automat. Control}
	\textbf{\bibinfo{volume}{58}}, \bibinfo{pages}{2205} (\bibinfo{year}{2013}).
	
	\bibitem[{\citenamefont{Boscain et~al.}(2012)\citenamefont{Boscain, Caponigro,
			Chambrion, and Sigalotti}}]{BCCS}
	\bibinfo{author}{\bibfnamefont{U.}~\bibnamefont{Boscain}},
	\bibinfo{author}{\bibfnamefont{M.}~\bibnamefont{Caponigro}},
	\bibinfo{author}{\bibfnamefont{T.}~\bibnamefont{Chambrion}},
	\bibnamefont{and}
	\bibinfo{author}{\bibfnamefont{M.}~\bibnamefont{Sigalotti}},
	\bibinfo{journal}{Comm. Math. Phys.} \textbf{\bibinfo{volume}{311}},
	\bibinfo{pages}{423} (\bibinfo{year}{2012}). 
	
	\bibitem[{\citenamefont{Boscain et~al.}(2014)\citenamefont{Boscain, Caponigro,
			and Sigalotti}}]{BCS}
	\bibinfo{author}{\bibfnamefont{U.}~\bibnamefont{Boscain}},
	\bibinfo{author}{\bibfnamefont{M.}~\bibnamefont{Caponigro}},
	\bibnamefont{and}
	\bibinfo{author}{\bibfnamefont{M.}~\bibnamefont{Sigalotti}},
	\bibinfo{journal}{J. Differential Equations} \textbf{\bibinfo{volume}{256}},
	\bibinfo{pages}{3524} (\bibinfo{year}{2014}).
	
	\bibitem[{\citenamefont{Boscain et~al.}(2021)\citenamefont{Boscain, Pozzoli,
			and Sigalotti}}]{Boscain19}
	\bibinfo{author}{\bibfnamefont{U.}~\bibnamefont{Boscain}},
	\bibinfo{author}{\bibfnamefont{E.}~\bibnamefont{Pozzoli}}, \bibnamefont{and}
	\bibinfo{author}{\bibfnamefont{M.}~\bibnamefont{Sigalotti}},
	\bibinfo{journal}{SIAM J. Control Optim.} \textbf{\bibinfo{volume}{59}},
	\bibinfo{pages}{156} (\bibinfo{year}{2021}).
	
	\bibitem[{\citenamefont{Leibscher et~al.}(2020)\citenamefont{Leibscher,
			Pozzoli, Perez, Schnell, Sigalotti, Boscain, and Koch}}]{Leibscher20}
	\bibinfo{author}{\bibfnamefont{M.}~\bibnamefont{Leibscher}},
	\bibinfo{author}{\bibfnamefont{E.}~\bibnamefont{Pozzoli}},
	\bibinfo{author}{\bibfnamefont{C.}~\bibnamefont{Perez}},
	\bibinfo{author}{\bibfnamefont{M.}~\bibnamefont{Schnell}},
	\bibinfo{author}{\bibfnamefont{M.}~\bibnamefont{Sigalotti}},
	\bibinfo{author}{\bibfnamefont{U.}~\bibnamefont{Boscain}}, \bibnamefont{and}
	\bibinfo{author}{\bibfnamefont{C.~P.} \bibnamefont{Koch}},
	\bibinfo{journal}{arXiv:2010.09296}  (\bibinfo{year}{2020}).
	
	\bibitem[{\citenamefont{Eibenberger et~al.}(2017)\citenamefont{Eibenberger,
			Doyle, and Patterson}}]{EibenbergerPRL17}
	\bibinfo{author}{\bibfnamefont{S.}~\bibnamefont{Eibenberger}},
	\bibinfo{author}{\bibfnamefont{J.}~\bibnamefont{Doyle}}, \bibnamefont{and}
	\bibinfo{author}{\bibfnamefont{D.}~\bibnamefont{Patterson}},
	\bibinfo{journal}{Phys. Rev. Lett.} \textbf{\bibinfo{volume}{118}},
	\bibinfo{pages}{123002} (\bibinfo{year}{2017}).
	
	\bibitem[{\citenamefont{P{\'e}rez et~al.}(2017)\citenamefont{P{\'e}rez, Steber,
			Domingos, Krin, Schmitz, and Schnell}}]{PerezAngewandte17}
	\bibinfo{author}{\bibfnamefont{C.}~\bibnamefont{P{\'e}rez}},
	\bibinfo{author}{\bibfnamefont{A.~L.} \bibnamefont{Steber}},
	\bibinfo{author}{\bibfnamefont{S.~R.} \bibnamefont{Domingos}},
	\bibinfo{author}{\bibfnamefont{A.}~\bibnamefont{Krin}},
	\bibinfo{author}{\bibfnamefont{D.}~\bibnamefont{Schmitz}}, \bibnamefont{and}
	\bibinfo{author}{\bibfnamefont{M.}~\bibnamefont{Schnell}},
	\bibinfo{journal}{Angew. Chem. Int. Ed.} \textbf{\bibinfo{volume}{56}},
	\bibinfo{pages}{12512} (\bibinfo{year}{2017}).
	
	\bibitem[{\citenamefont{P{\'e}rez et~al.}(2018)\citenamefont{P{\'e}rez, Steber,
			Krin, and Schnell}}]{PerezJPCL18}
	\bibinfo{author}{\bibfnamefont{C.}~\bibnamefont{P{\'e}rez}},
	\bibinfo{author}{\bibfnamefont{A.~L.} \bibnamefont{Steber}},
	\bibinfo{author}{\bibfnamefont{A.}~\bibnamefont{Krin}}, \bibnamefont{and}
	\bibinfo{author}{\bibfnamefont{M.}~\bibnamefont{Schnell}},
	\bibinfo{journal}{J. Phys. Chem. Lett.} \textbf{\bibinfo{volume}{9}},
	\bibinfo{pages}{4539} (\bibinfo{year}{2018}).
	
	\bibitem[{\citenamefont{Zare}(1988)}]{Zare88}
	\bibinfo{author}{\bibfnamefont{R.~N.} \bibnamefont{Zare}},
	\emph{\bibinfo{title}{Angular Momentum}} (\bibinfo{publisher}{Wiley},
	\bibinfo{year}{1988}).
	
	\bibitem[{\citenamefont{Koch et~al.}(2019)\citenamefont{Koch, Lemeshko, and
			Sugny}}]{KochRMP}
	\bibinfo{author}{\bibfnamefont{C.~P.} \bibnamefont{Koch}},
	\bibinfo{author}{\bibfnamefont{M.}~\bibnamefont{Lemeshko}}, \bibnamefont{and}
	\bibinfo{author}{\bibfnamefont{D.}~\bibnamefont{Sugny}},
	\bibinfo{journal}{Rev. Mod. Phys.} \textbf{\bibinfo{volume}{91}},
	\bibinfo{pages}{035005} (\bibinfo{year}{2019}).
	
	\bibitem[{\citenamefont{Abramowitz and (eds.)}(1964)}]{Abramowitz64}
	\bibinfo{author}{\bibfnamefont{M.}~\bibnamefont{Abramowitz}} \bibnamefont{and}
	\bibinfo{author}{\bibfnamefont{I.~A.~S.} \bibnamefont{(eds.)}},
	\emph{\bibinfo{title}{Handbook of mathematical functions}}
	(\bibinfo{publisher}{United States Department of Commerce, National Bureau of
		Standards}, \bibinfo{year}{1964}).
	
	\bibitem[{\citenamefont{Godsil and Severini}(2010)}]{GodsilPRA10}
	\bibinfo{author}{\bibfnamefont{C.}~\bibnamefont{Godsil}} \bibnamefont{and}
	\bibinfo{author}{\bibfnamefont{S.}~\bibnamefont{Severini}},
	\bibinfo{journal}{Phys. Rev. A} \textbf{\bibinfo{volume}{81}},
	\bibinfo{pages}{052316} (\bibinfo{year}{2010}).
	
	\bibitem[{\citenamefont{Gokler et~al.}(2017)\citenamefont{Gokler, Lloyd, Shor,
			and Thompson}}]{GoklerPRL17}
	\bibinfo{author}{\bibfnamefont{C.}~\bibnamefont{Gokler}},
	\bibinfo{author}{\bibfnamefont{S.}~\bibnamefont{Lloyd}},
	\bibinfo{author}{\bibfnamefont{P.}~\bibnamefont{Shor}}, \bibnamefont{and}
	\bibinfo{author}{\bibfnamefont{K.}~\bibnamefont{Thompson}},
	\bibinfo{journal}{Phys. Rev. Lett.} \textbf{\bibinfo{volume}{118}},
	\bibinfo{pages}{260501} (\bibinfo{year}{2017}).
	
	\bibitem[{\citenamefont{Pozzoli}(2021)}]{Pozzoli2021}
	\bibinfo{author}{\bibfnamefont{E.}~\bibnamefont{Pozzoli}},
	\bibinfo{journal}{arXiv:2108.01943}  (\bibinfo{year}{2021}).
	
	\bibitem[{\citenamefont{Schirmer et~al.}(2008)\citenamefont{Schirmer, Pullen,
			and Pemberton-Ross}}]{SchirmerPRA08}
	\bibinfo{author}{\bibfnamefont{S.~G.} \bibnamefont{Schirmer}},
	\bibinfo{author}{\bibfnamefont{I.~C.~H.} \bibnamefont{Pullen}},
	\bibnamefont{and} \bibinfo{author}{\bibfnamefont{P.~J.}
		\bibnamefont{Pemberton-Ross}}, \bibinfo{journal}{Phys. Rev. A}
	\textbf{\bibinfo{volume}{78}}, \bibinfo{pages}{062339}
	(\bibinfo{year}{2008}).
	
	\bibitem[{\citenamefont{Wang et~al.}(2012)\citenamefont{Wang, Pemberton-Ross,
			and Schirmer}}]{WangIEEE12}
	\bibinfo{author}{\bibfnamefont{X.}~\bibnamefont{Wang}},
	\bibinfo{author}{\bibfnamefont{P.}~\bibnamefont{Pemberton-Ross}},
	\bibnamefont{and} \bibinfo{author}{\bibfnamefont{S.~G.}
		\bibnamefont{Schirmer}}, \bibinfo{journal}{IEEE Transactions on Automatic
		Control} \textbf{\bibinfo{volume}{57}}, \bibinfo{pages}{1945}
	(\bibinfo{year}{2012}).
	
	\bibitem[{\citenamefont{Wang et~al.}(2016)\citenamefont{Wang, Burgarth, and
			Schirmer}}]{WangPRA16}
	\bibinfo{author}{\bibfnamefont{X.}~\bibnamefont{Wang}},
	\bibinfo{author}{\bibfnamefont{D.}~\bibnamefont{Burgarth}}, \bibnamefont{and}
	\bibinfo{author}{\bibfnamefont{S.}~\bibnamefont{Schirmer}},
	\bibinfo{journal}{Phys. Rev. A} \textbf{\bibinfo{volume}{94}},
	\bibinfo{pages}{052319} (\bibinfo{year}{2016}).
	
	\bibitem[{\citenamefont{Chen et~al.}(2020)\citenamefont{Chen, Zhou, Bian, Li,
			and Peng}}]{ChenPRA20}
	\bibinfo{author}{\bibfnamefont{J.}~\bibnamefont{Chen}},
	\bibinfo{author}{\bibfnamefont{Y.}~\bibnamefont{Zhou}},
	\bibinfo{author}{\bibfnamefont{J.}~\bibnamefont{Bian}},
	\bibinfo{author}{\bibfnamefont{J.}~\bibnamefont{Li}}, \bibnamefont{and}
	\bibinfo{author}{\bibfnamefont{X.}~\bibnamefont{Peng}},
	\bibinfo{journal}{Phys. Rev. A} \textbf{\bibinfo{volume}{102}},
	\bibinfo{pages}{032602} (\bibinfo{year}{2020}).
	
	\bibitem[{\citenamefont{Albertini and D’Alessandro}(2021)}]{Albertini21}
	\bibinfo{author}{\bibfnamefont{F.}~\bibnamefont{Albertini}} \bibnamefont{and}
	\bibinfo{author}{\bibfnamefont{D.}~\bibnamefont{D’Alessandro}},
	\bibinfo{journal}{Systems \& Control Letters} \textbf{\bibinfo{volume}{151}},
	\bibinfo{pages}{104913} (\bibinfo{year}{2021}).
	
\end{thebibliography}

\end{document}